\documentclass[%
reprint,
twocolumn,
superscriptaddress,
showpacs,preprintnumbers,
nofootinbib,
aps,
prd,
]{revtex4-2}
\usepackage{amsfonts}
\usepackage{bbm,bm}
\usepackage[breaklinks,urlbordercolor={1 1 1}]{hyperref}
\usepackage{graphicx}
\usepackage{amsmath}
\usepackage{amssymb}
\usepackage{mathtools}
\usepackage{hyperref}
\usepackage{graphicx}
\usepackage{xcolor}
\usepackage{comment}
\usepackage{subfigure}
\usepackage{dcolumn}
\usepackage[utf8]{inputenc} 
\usepackage{multirow}
\usepackage{float}
\usepackage{booktabs}

\usepackage{ulem,fancyvrb}
\usepackage{xcolor}

\newcommand{\be}{\begin{equation}} 
\newcommand{\ee}{\end{equation}}
\newcommand{\nn }{\nonumber        }
\newcommand{\ba}{\begin{eqnarray}}
\newcommand{\ea}{\end{eqnarray}}
\newcommand{\abs}[1]{\left\lvert #1\right\rvert}

\begin{document}

\newcommand{\ITP}{\affiliation{CAS Key Laboratory of Theoretical Physics, Institute of Theoretical Physics,\\
Chinese Academy of Sciences, Beijing 100190, China}}

\newcommand{\ICTPA}{\affiliation{International Centre for Theoretical Physics Asia-Pacific, Beijing/Hangzhou, China}}

\newcommand{\HIAS}{\affiliation{School of Fundamental Physics and Mathematical Sciences, Hangzhou Institute for Advanced
Study, UCAS, Hangzhou 310024, China}}

\newcommand{\PKU}{\affiliation{Center for High Energy Physics, Peking University, Beijing 100871, China}}

\newcommand{\UCAS}{\affiliation{School of Physical Sciences, University of Chinese Academy of Sciences, Beijing 100049, P.\ R.\ China}}

\newcommand{\TSUa}{\affiliation{Tsung-Dao Lee Institute \& School of Physics and Astronomy, Shanghai Jiao Tong University, Shanghai 200240, China }}

\newcommand{\TSUb}{\affiliation{Shanghai Key Laboratory for Particle Physics and Cosmology, Key Laboratory for Particle Astrophysics and Cosmology (MOE), Shanghai Jiao Tong University, Shanghai 200240, China }}

\newcommand{\AMH}{\affiliation{Amherst Center for Fundamental Interactions, Department of Physics,\\
University of Massachusetts, Amherst,
MA 01003, USA }}

\newcommand{\CAL}{\affiliation{Kellogg Radiation Laboratory, California Institute of Technology,\\
Pasadena,
CA 91125, USA}}
\newcommand{\PU}{\affiliation{Faculty of Fundamental Sciences, PHENIKAA University, Yen Nghia, Ha Dong, Hanoi 12116, Vietnam}}

\newcommand{\UCLouvain}{\affiliation{Centre for Cosmology, Particle Physics and Phenomenology, Universit\'{e}~catholique~de~Louvain, Louvain-la-Neuve B-1348, Belgium}}


\hfill ACFI T24-20

\title{Does the Electron EDM Preclude Electroweak Baryogenesis ?}

\author{Yuan-Zhen Li} \email{liyuanzhen@itp.ac.cn, yuanzhen.li@uclouvain.be} \ITP \UCAS \UCLouvain

\author{Michael J.~Ramsey-Musolf} \email{mjrm@sjtu.edu.cn, mjrm@physics.umass.edu} \TSUa \TSUb \AMH \CAL

\author{Jiang-Hao Yu} \email{jhyu@itp.ac.cn}  \ITP \UCAS \PKU \HIAS \ICTPA

\begin{abstract}

Electroweak baryogenesis (EWBG) constitutes a theoretically compelling and experimentally testable mechanism for explaining the origin of the baryon asymmetry of the universe (BAU). New results for the electric dipole moment (EDM) of the electron place significant constraints on the beyond Standard Model CP-violation needed for successful EWBG. {Using a specific model illustration,} we show how new developments in EWBG quantum transport theory that include CP-violating sources first order in gradients imply more relaxed EDM constraints {-- and thereby greater EWBG viability -- }than implied by previous approximation formulations. 
We also illustrate how these developments enable a more realistic treatment of CP-conserving interactions that can also have a decisive impact on the predicted BAU.

\end{abstract}

\maketitle

Explaining the origin of the baryon asymmetry of the universe (BAU) is a key unsolved problem at the interface of particle and nuclear physics with cosmology. Both the mechanism for baryogenesis as well as the early universe era in which it occurred remain unknown. A compelling possibility is electroweak baryogenesis (EWBG), which links the BAU to the spontaneous electroweak symmetry breaking (EWSB) 
and generation of elementary particle masses via the Higgs mechanism.\cite{Kuzmin:1985mm,Shaposhnikov:1987pf,Shaposhnikov:1987tw} (For reviews, see {\it e.g.},  \cite{Trodden:1998ym,Cline:2006ts,Morrissey:2012db}. ) In principle, the Standard Model (SM) of particle physics contains the necessary ingredients for EWBG \cite{Sakharov:1967dj}: B-violation via electroweak (EW) sphaleron processes; C- and CP-violation in the electroweak sector; and out-of-equilibrium conditions in the guise of a first order electroweak phase transition (FOEWPT) to the present Higgs phase. In practice, the latter does not occur for a  Higgs boson heavier than $\sim 70-80$ GeV \cite{Bochkarev:1987wf,Kajantie:1995kf, Laine:2012jy}, while the effects of CP-violation  {(CPV)} in the Cabibbo-Kobayashi-Maskawa (CKM) matrix are too feeble to have generated the observed BAU, even for a sufficiently light Higgs boson \cite{Gavela:1993ts,Huet:1994jb,Gavela:1994dt}.

Physics beyond the Standard Model (BSM) can remedy these shortcomings. 
An extended scalar sector can readily lead to a FOEWPT even for a 125 GeV Higgs boson (see \cite{Ramsey-Musolf:2019lsf} for  extensive  of references), while providing the efficient CPV.
The requisite mass scale for these new particles ($\lesssim 700$ GeV) as well as the needed strength of their coupling to the Higgs boson generically puts them within the reach of future high energy collider searches and precision Higgs boson studies\cite{Ramsey-Musolf:2019lsf}. Results from the Large Hadron Collider do not preclude such an extended scalar sector, and it may require a future 100 TeV $pp$ collider to provide a definitive test \cite{Ramsey-Musolf:2019lsf}. 
Next generation gravitational wave detectors, such as LISA, Taiji, and Tianqin, provide a complementary probe and could uncover a stochastic gravitational wave background arising from a FOEWPT \cite{Caprini:2015zlo,Caprini:2019egz,Crowder:2005nr}.

Searches for the permanent electric dipole moments (EDMs) of atoms, molecules, and nucleons provide the most powerful probe of the BSM CPV needed for EWBG \cite{Chupp:2014gka,Engel:2013lsa,Pospelov:2005pr}. 
Theoretically, drawing 
quantitative inferences about EWBG viability from EDM search results
requires performing robust computations of the early universe CPV dynamics. Here, we report on advances addressing this challenge and the corresponding implications for the EDM-EWBG connection. 

The EWBG CPV dynamics occur during a FOEWPT that proceeds via nucleation of bubbles of broken electroweak symmetry, defined by regions of non-vanishing, spacetime varying scalar background fields $\varphi(x)$ ({\it i.e.}, the Higgs field). 
CPV-interactions at the bubble walls induce a non-zero density of left-handed SM fermions, $n_L$, that diffuses into the symmetric phase, biasing EW sphaleron transitions into creation of a net B$+$L number. The latter diffuses back inside the expanding bubbles, where EWSB quenches the sphalerons and preserves the BAU, assuming a sufficiently strong FOEWPT. 

The challenge in computing $n_L$
entails solving -- in the presence of  $\varphi(x)$ -- the quantum transport equations for Greens functions that encode information on particle densities. The mass of any particle that interacts with the $\varphi(x)$ varies with spacetime as it traverses the bubble wall, necessitating a continual re-definition of the mass eigenstates. Previous EWBG computations have employed various approaches to solving these transport dynamics\cite{Cline:1997vk,Konstandin:2013caa,Garbrecht:2018mrp,Bell:2019mbn,Basler:2021kgq,Kainulainen:2021oqs,Cline:2021dkf,Cline:2021iff,Carena:2018cjh,Carena:2019xrr,Carena:2022qpf,Enomoto:2021dkl}.
For a given set of CPV parameters, the resulting BAU predictions can vary by an order of magnitude. The most optimistic typically result from the use of the \lq\lq vev insertion approximation\rq\rq\, (VIA) \cite{Riotto:1995hh,Riotto:1997vy,Lee:2004we, Cirigliano:2006wh,Postma:2019scv}, whose theoretical consistency has been criticized recently in Refs.~\cite{Cline:2020jre,Kainulainen:2021oqs,Postma:2022dbr}. In the proposed alternative, semiclassical (SC) formulation \cite{Cline:2020jre,Kainulainen:2021oqs}, the CPV source terms first arise at second order in gradients with respect to position along the bubble wall profile, leading to a significantly smaller BAU than in the VIA (for a review, see Ref.~\cite{Garbrecht:2018mrp}). 
{As a result, the corresponding viability of EWBG with given EDM limits is significantly suppressed or even precluded, according to the predictions of the SC method. The robustness of the transport theory is therefore a key ingredient for the EWBG-EDM interface.}

{In what follows, we provide a consistent treatment of the EWBG CPV dynamics including the first-order-in-gradients CPV sources \cite{Cirigliano:2009yt,Cirigliano:2011di}, as well as robust CP-conserving interactions, which avoids both the VIA inconsistencies and the SC approximations.
We demonstrate that despite its theoretical shortcomings, the VIA as employed in earlier work can under-predict the magnitude of the BAU, in contrast to the conclusions drawn from the SC treatments. This
enhancement in BAU arises from a consistent treatment of flavor
mixing in the presence of a spacetime-varying background field, a feature missed by the SC expansion \cite{Cline:2020jre,Kainulainen:2021oqs}. Consequently, the viability of the EWBG model is significantly enhanced, {thereby enlarging the EWBG-consistent parameter space in a given model}. }

Employing a realistic  EWBG model \cite{Patel:2012pi,Inoue:2015pza} for illustration, we solve the Kadanoff-Baym transport equations \cite{Schwinger:1960qe, Mahanthappa:1962ex, Bakshi:1962dv, Bakshi:1963bn, Keldysh:1964ud,Chou:1984es} using the vev resummation (VR) framework developed in Refs.~\cite{Cirigliano:2009yt,Cirigliano:2011di}. 
(See~\cite{Cirigliano:2011di} for a detailed delineation of differences between the VR and SC frameworks.) 
For a given set of model parameters, the VR result for the BAU can be as large or even a few times larger than the VIA prediction. Consequently, EDM constraints on EWBG can be more relaxed than previously realized. 
We also provide a realistic, quantitative determination of the dependence of the BAU transport dynamics on model parameters -- including those that enter the CP-conserving \lq\lq collision terms\rq\rq\, -- a feature that has typically eluded earlier studies. While there remain open challenges pertaining to bubble wall dynamics \cite{Dine:1992wr,Moore:2000wx,Espinosa:2010hh,Bodeker:2017cim,Hoche:2020ysm,DeCurtis:2022hlx,Laurent:2022jrs},  the results reported herein constitute a significant advance for assessing the EWBG-EDM interface. 
{We stress that while we illustrate these effects within a specific scalar extension of the Standard Model, the conclusions are generic for any model involving background field-induced CPV mixing between two species, including those with fermionic sources. In this context, the scalar model we {adopt} serves as a {paradigmatic case study} for assessing the EWBG-EDM interface. The resulting relaxation of EDM constraints {underscores the importance of} EWBG as a continuing primary motivation for the worldwide EDM experimental program.
}

We introduce general features of the scalar field transport dynamics {(see \cite{Inoue:2015pza})} before describing the model illustration. Consider a model with two complex, electrically neutral scalar fields $H^0_{1,2}$ denoted  by the \lq\lq flavor space\rq\rq\,  vector $\eta\equiv (H^0_1, H^0_2)$. The two flavor components of $\eta$ interact with scalar fields ${\hat\phi}_k(x)$, whose classical values $\varphi_k(x)$ define the bubble walls. The $\eta$-$\varphi_k(x)$ interactions lead to a 
mass-squared matrix having the generic form
\begin{equation}
M^2_\eta(x) = \left( 
\begin{array}{cc} 
M_1^2(x) & R(x) \, e^{-i \, \alpha(x)} \\ R(x) \, e^{i \, \alpha(x)} & M_2^2(x) 
\end{array}
\right) \; , \label{eq:wow}
\end{equation}
where $M_{1,2}(x), \, R(x), \, \alpha(x)$ depend on the model parameters and the spacetime-dependence of the $\varphi_k(x)$.

We solve for the neutral scalar Greens functions  by first diagonalizing $M_\eta^2(x)$ at each spacetime point using a unitarity transformation ${\hat \eta}=U(x) {\eta}$, where the hatted fields correspond to the mass eigenstates with diagonal mass-squared matrix ${\hat m}^2(x)$. Evolution of the mass basis particle (anti-particle) density matrices $f_m$ ($\bar f_m$) follows from Schwinger-Dyson (SD) equations  for the 
scalar field Wightman functions $G_{ij}^<(x,y) \equiv \langle {\hat H}_j^\dag(y) {\hat H}_i(x)\rangle$ and $G_{ij}^>(x,y) \equiv \langle {\hat H}_i(x) {\hat H}_j^\dag (y)\rangle$, where $i,j\in\{1,2\}$. Following\cite{Cirigliano:2009yt,Cirigliano:2011di}, we transform to Wigner space co-ordinates $X=(x+y)/2$ and $k$, the wavenumber associated with the relative co-ordinate $x-y$, 
and reorganize the corresponding SD equations into the Kadanoff-Baym (KB) constraint and kinetic equations.

Observing that there exists a hierarchy of length scales in the problem facilitates a tractable solution to the KB equations. We define the  scale ratios: $\epsilon_w \equiv {L_\mathrm{int}}/{L_w}$ , $\epsilon_\mathrm{coll} \equiv {L_\mathrm{int}}/{L_\mathrm{mfp}}$, and $\epsilon_\mathrm{osc} \equiv {L_\mathrm{int}}/{L_\mathrm{osc}}$, where $L_\mathrm{int} = |{\vec k}|\sim T^{-1}$ is the de Broglie wavelength with $T$ being the temperture of the plasma; $L_w$ is the wall thickness, which in many models is $\mathrm{O}(10/T) $, so that $L_w \gg L_\mathrm{int}$; $L_\mathrm{osc}$ is the length scale associated with " flavor" oscillations $H_1^0\leftrightarrow H_2^0$; and $L_\mathrm{mfp}$ is the mean free path associated with gauge and scalar field interactions. For the scenarios of interest here, one finds $L_\mathrm{mfp} \gg L_\mathrm{int}$ for perturbative values of the couplings, while  {CPV asymmetries are maximized for $L_w \sim L_\mathrm{osc}$} in the " thick wall" regime\cite{Cirigliano:2009yt}.
Thus, one has $\epsilon_{w,\mathrm{coll},\mathrm{osc}} \ll 1$  {in the interesting region}.

Expanding the constraint and kinetic equations to orders $\epsilon^0$ and $\epsilon$, respectively, yields the following quantum Boltzmann equations for the density matrices:
\begin{subequations}
\begin{align}
\nn
(u\cdot \partial_X + {\vec F}\cdot\nabla_k ) f_m({\vec k},X) & = -\left[ i\omega_k +u\cdot\Sigma, f_m({\vec k},X)\right]\\
\label{eq:qb1}
&+\mathcal{C}_m[f_m, {\bar f_m}]({\vec k},X)\\
\nn
(u\cdot \partial_X + {\vec F}\cdot\nabla_k ) {\bar f}_m({\vec k},X) & = +\left[ i\omega_k - u\cdot\Sigma, {\bar f}_m({\vec k},X)\right]\\
\label{eq:qb2}
&+\mathcal{C}_m[ {\bar f_m}, f_m]({\vec k},X)\ \ \ ,
\end{align}
\end{subequations}
where $u^\mu=(1,{\vec v}) ; \, {\vec v} = {\vec k}/{\bar\omega}_k  ; \, {\bar\omega}_k  = \sqrt{|{\vec k}|^2+{\bar m}^2(x)} $; ${\bar m}^2=(M_1^2+M^2_2)/2 ; \; {\vec F}  = \nabla_X {\bar\omega}_k$;  {$\omega_k = {\rm diag} \{ \omega_{1k}, \omega_{2k}\}$; $\omega_{ik} =  \sqrt{|{\vec k}|^2+{ M_i}^2(x)}$;} $\Sigma^\mu = U^\dag \partial^\mu U$, and the \lq\lq collision term" $\mathcal{C}_m$ is a functional of the $f_m$ and ${\bar f_m}$.   

Note that the terms in the LHS of (\ref{eq:qb1},\ref{eq:qb2}) generalize the space-time derivative and force terms in classical Boltzmann equation. The \lq\lq force\rq\rq\, $ {\vec F} $ is associated with the variation of the background fields which contribute to $\bar m(x)$.  On the RHS, the commutator $- i[\omega_k, f_m]$  ($- i[\omega_k, {\bar f}_m]$) gives rise to (anti-)particle flavor oscillations and is identical in form to what appears in the familiar density matrix formalism for neutrino flavor oscillations. 
{Crucially, the commutators $[u\cdot\Sigma, f_m]$ and $[u\cdot\Sigma, {\bar f}_m]$ encode the CP-violating sources. From the definition of $\Sigma^\mu \equiv U^\dag \partial^\mu U$, it is evident that these sources are linear in the spacetime gradients of the background fields ($\partial^\mu \theta, \partial^\mu \alpha$). This stands in contrast to the semiclassical (SC) approximation, where CPV sources typically appear at second order in gradients. }
 Finally, the collision term $\mathcal{C}_m$ embodies the effect of all interactions that lead to thermalization in the plasma, chemical equilibrium associated with particle species changing reactions, and diffusion ahead of the advancing bubble wall. 

{The relative sign difference between the oscillation term and CPV source terms in Eqs.~(\ref{eq:qb1}, \ref{eq:qb2}) 
leads to a net number density (a.k.a., CPV asymmetry) for a given particle species. Writing 
\begin{equation}
U= \left(
\begin{array}{cc}
\cos\theta(x) & -\sin\theta(x)e^{-i\alpha(x)}\\
\sin\theta(x)e^{i\alpha(x)} & \cos\theta(x)
\end{array}
\right)
\end{equation}
}
{gives $\Sigma^\mu=$
\begin{equation}
\left(
\begin{array}{cc}
0 & \ e^{-i\alpha}\\
e^{i\alpha} & 0
\end{array}
\right)\partial^\mu\theta
+\frac{i}{2}\left(
\begin{array}{cc}
2\sin^2\theta & \ \sin{2\theta} e^{-i\alpha}\\
\sin{2\theta}e^{i\alpha} & -2\sin^2\theta
\end{array}
\right)\partial^\mu\alpha
\nn.
\end{equation}
}
{Combining the first two terms on the RHS of Eqs. (\ref{eq:qb2},\ref{eq:qb2}) leads, {\it e.g.}, for the $(1,1)$ element 
\begin{subequations}
\begin{align}
\left(u\cdot\partial_X + {\vec F}\cdot{\vec\nabla}\right)f_{11} &\supset -\left[\, i(\omega_k+\sin^2\theta\, u\cdot\partial\alpha), f\,\right]_{11}+\cdots 
\nn\\
\left(u\cdot\partial_X + {\vec F}\cdot{\vec\nabla}\right){\bar f}_{11} &\supset \left[\, i(\omega_k-\sin^2\theta\,  u\cdot\partial\alpha), {\bar f}\, \right]_{11}+\cdots
\nn
\end{align}
\end{subequations}
The simultaneous presence of flavor mixing ($\sin\theta\not=0$) and a spacetime dependent phase ($\partial^\mu\alpha\not=0$) thus leads to a difference in the particle and antiparticle oscillation frequencies proportional to $\sin^2\theta u\cdot\partial \alpha$ and to a corresponding, non-vanishing CPV asymmetry, $f_{11}-\bar{f}_{11}$.
}

{To illustrate the impact of this effect using the VR framework,} we  solve Eqs.~(\ref{eq:qb1},\ref{eq:qb2}) for the \lq\lq Two-Step EWBG\rq\rq\, model of Refs.~\cite{Patel:2012pi,Inoue:2015pza}. 
Baryogenesis occurs during the first of two successive electroweak symmetry-breaking (EWSB) transitions, wherein the $\varphi_k(x)\not=0$ while the components of $\eta$ admit no non-vanishing background field values. For renormalizable $\eta$-${\hat\phi}_k$ interactions,  the emergence of a spacetime varying phase $\alpha(x)$ in Eq.~(\ref{eq:wow}) during the first step requires the presence of at least two non-vanishing $\varphi_k(x)$ \cite{Inoue:2015pza}. Thus, one requires at least four scalar fields: the two components of $\eta$ and the two ${\hat\phi}_k$. 

A minimal realization entails a scalar sector consisting of two Higgs doublets $H_{1,2}$, a hypercharge $Y=0$ real triplet $\Sigma$, and a SM gauge singlet $S$. All scalars are SU(3)$_C$ singlets. The gauge and fermion sectors are unchanged from the SM. To model the impact of the latter, we introduce an additional scalar field $A$, whose dynamics implement all other flavor-diagonal thermalizing interactions in the plasma, such as those arising from gauge and Yukawa interactions. During the first EWSB transition, $S$ and the neutral component of $\Sigma$ obtain vacuum expectation values (vevs), $v_s$ and $v_\sigma$, respectively, with corresponding field flucuations described by $s=S-v_s$ and $\sigma = \Sigma^0-v_\sigma$. These vevs vary with spacetime, thereby providing the requisite two background fields $\varphi_k(x)$ with $k=1,2$. In the second transition, ($v_s$, $v_\sigma$) relax to zero while the neutral components of the doublets obtain vevs, $v_{1,2}$, with $\sqrt{v_1^2+v_2^2}=246$ GeV. One may embed the model in a supersymmetric context\cite{Bandyopadhyay:2015tva,Bandyopadhyay:2015oga}, with the corresponding superpartners augmenting the field content. Here we consider the non-supersymmetric version.

For successful EWBG during the first step, this transition must be first order, a condition shown to be satisfied in both perturbative and non-perturbative (lattice) computations for suitable choices of the scalar potential parameters \cite{Patel:2011th,Niemi:2020hto}.
CPV interactions between the $H_{1,2}$ and the $(S,\Sigma)$ vevs catalyze generation of non-zero Higgs number densities, $n_{H_{1,2}}$. Yukawa interactions then transfer the latter into non-vanishing fermion number densities. {The resultant left-handed fermion densities} bias electroweak sphalerons into producing a non-zero B$+$L density that diffuses into the bubble interiors. 


The scalar potential is  $V(H_1,H_2,\Sigma,S,A) = V_H+V_\phi+V_{H\phi}$, where $V_H$ is the CP-conserving Two Higgs Double Model (2HDM) potential \cite{Gunion:2002zf,Branco:2011iw,Inoue:2014nva}, $V_\phi$ involves only the $\phi\equiv (S,\Sigma,A)$ fields, and the key \lq\lq portal\rq\rq\, interaction terms are contained in
\begin{eqnarray}
\nonumber
V_{H\phi}\supset \frac{1}{2} H_1^\dag H_2\left(a_1 S^2 + a_2 \Sigma^2\right)+ \mathrm{h.c.}\\
+ \sum_{i=1,2}\left[y_1^{ii}S^2+y_2^{ii}\Sigma^2+y_3^{ii}A^2\right]H_i^\dag H_i\ \ \ .
\label{eq:vint}
\end{eqnarray}
The physical (rephasing-invariant) CPV phases are $\delta_S=\mathrm{arg}(a_1^\ast v_1 v_2^\ast)$ and $\delta_\Sigma=\mathrm{arg}(a_2^\ast v_1 v_2^\ast)$. A combination of these CPV phases and the $(S,\Sigma)$ vevs induce the $M_i^2(x)$ as well as the $\alpha(x)$ in Eq.~(\ref{eq:wow}) and, thus, the CPV sources in the KB equations{, as seen in the Supplementary Material}. The interactions in Eq.~(\ref{eq:vint}) also give rise to Higgs flavor off-diagonal collision terms, which we include in the computation. 
The $A$ fields do not obtain vacuum expectation values and, thus, do not contribute to the spacetime-dependence in $M_\eta^2(x)$. 

To solve Eqs.~(\ref{eq:qb1}, \ref{eq:qb2})
{we choose the couplings $y_a^{ii}$ ($a=1,2$, $i=1,2$) so as to yield ${\bar m}^2$ $x$-independent, implying that} ${\vec F}=0$ in our set up. {Doing so allows a direct comparison with the results in Ref.~\cite{Cirigliano:2011di}; we will investigate the impact of ${\vec F}\not=0$ in future work.} 
We then make additional simplifying assumptions relevant to the collision integrals $\mathcal{C}_m[f_m, {\bar f_m}]$ outlined in \cite{Cirigliano:2011di,Inoue:2015pza}. 
We also consider a type I 2HDM in which only one of the Higgs doublets has Yukawa interactions with the third generation up-type quarks that are in chemical equilibrium. For the interactions of the Higgs particles with the fields $A$, we assume the corresponding rates  $\Gamma_A$ are large (small) compared to the weak sphaleron and Yukawa (strong sphaleron) interaction rates. We thus obtain $n_L=(4 c_T+ 5c_Q) n_{H_1}$, where $c_{T,Q}$ are functions of statistical factors $k_j$ relating the number density for a given species $n_j$ to its chemical potential $\mu_j$.  {In addition, consider planar bubble walls so that physical quantities depend only on the comoving coordinate $z = X+ v_w t$, the distance to the wall,  with $v_w$ the wall velocity.}

The Higgs number density $n_{H_1}$ is obtained by (i) solving the quantum Boltzmann equations (\ref{eq:qb1},\ref{eq:qb2});  
{(ii) integrating the difference of mass-basis densities matrices to obtain the mass-basis number density ${\hat n}$};
and (iii) {inverting $\eta=U^\dag {\hat\eta}$  to obtain the flavor basis density for $H_1$ as} 
$
n_{H_1} = \left[ U(X)\, {\hat n}(X)\, U(X)^\dag\right]_{11}$.
The baryon number density is then
given by
\begin{equation}\label{eq:nB}
n_B=-3 \frac{\Gamma_{\mathrm{ws}}}{v_w} \int^0_{-\infty} d z\,  n_L(z) \exp \left(\frac{15}{4} \frac{\Gamma_{\mathrm{ws}}}{v_w} z\right),
\end{equation}
where 
we have integrated over the region of
unbroken EW symmetry in which $\Gamma_{\mathrm{ws}} $ is unsuppressed.

To obtain a numerical solution to Eqs.~(\ref{eq:qb1},\ref{eq:qb2}), which comprise a system of eight
coupled integro-differential equations (the ${f}_m$ and ${{\bar f}}_m$ are $2\times 2$ matrices in the  mass basis), 
we discretize $k$ and $\cos \theta_k$ into $N_k$ and $N_\theta$ bins within the ranges $ 0<k<k_{\rm max},\quad -1< \cos \theta_k <1 $
and take the central values of each bin. 
The  Boltzmann equations then yield a system of $ 8 \times N_k \times N_\theta$ coupled first order ordinary differential equations with boundary conditions, which we solve with the \lq\lq relaxation method\rq\rq \cite{Press:1992zz}. 
{Far from the wall, the collision terms bring the density matrices to their equilibrium forms in the positive time direction. Thus, we 
impose thermal-equilibrium boundary conditions in the negative (positive) time directions for the right-moving (left-moving) modes.}

We will compare our results to those obtained in the VIA. The latter framework treats the $(S,\Sigma^0)$ vevs as perturbative insertions, and otherwise utilizes flavor basis Greens functions. Note that flavor non-diagonal collision terms arising from interactions between the $H_{1,2}$ with $s$ and $\sigma$ and arising from the first line in Eq.~(\ref{eq:vint}), are absent in the VIA treatment.  
When comparing our results with those of the VIA computation, we follow the methods used in Ref.~\cite{Inoue:2015pza}. 
For 
the $v_s(x)$ and $v_\sigma(x)$ profiles we adopt the forms in Ref.~\cite{Inoue:2015pza}, along with the corresponding profile parameter values as well as wall velocity, $v_w=0.05$. 
The benchmark parameter choices are {given in the Supplementary Material}.

Fig.~\ref{fig:nH_nL} shows the resulting VR and VIA profiles $n_L(z)$ as a function of the distance normal to the bubble wall. The pronounced structure near $z=0$ reflects the variation in the bubble profiles near the wall center and the corresponding impact on the CPV sources involving $u\cdot\Sigma$ entering the RHS of Eqs.~(\ref{eq:qb1},\ref{eq:qb2}). Importantly, the VR diffusion tail ($z<0$) is significantly enhanced as compared to the VIA result. As the resulting value of $n_B$ entails integrating over this tail as in Eq.~(\ref{eq:nB}) we expect the VR to yield a larger baryon asymmetry. 

\begin{figure}[ht]
\centering
     \includegraphics[width = 0.9 \linewidth]{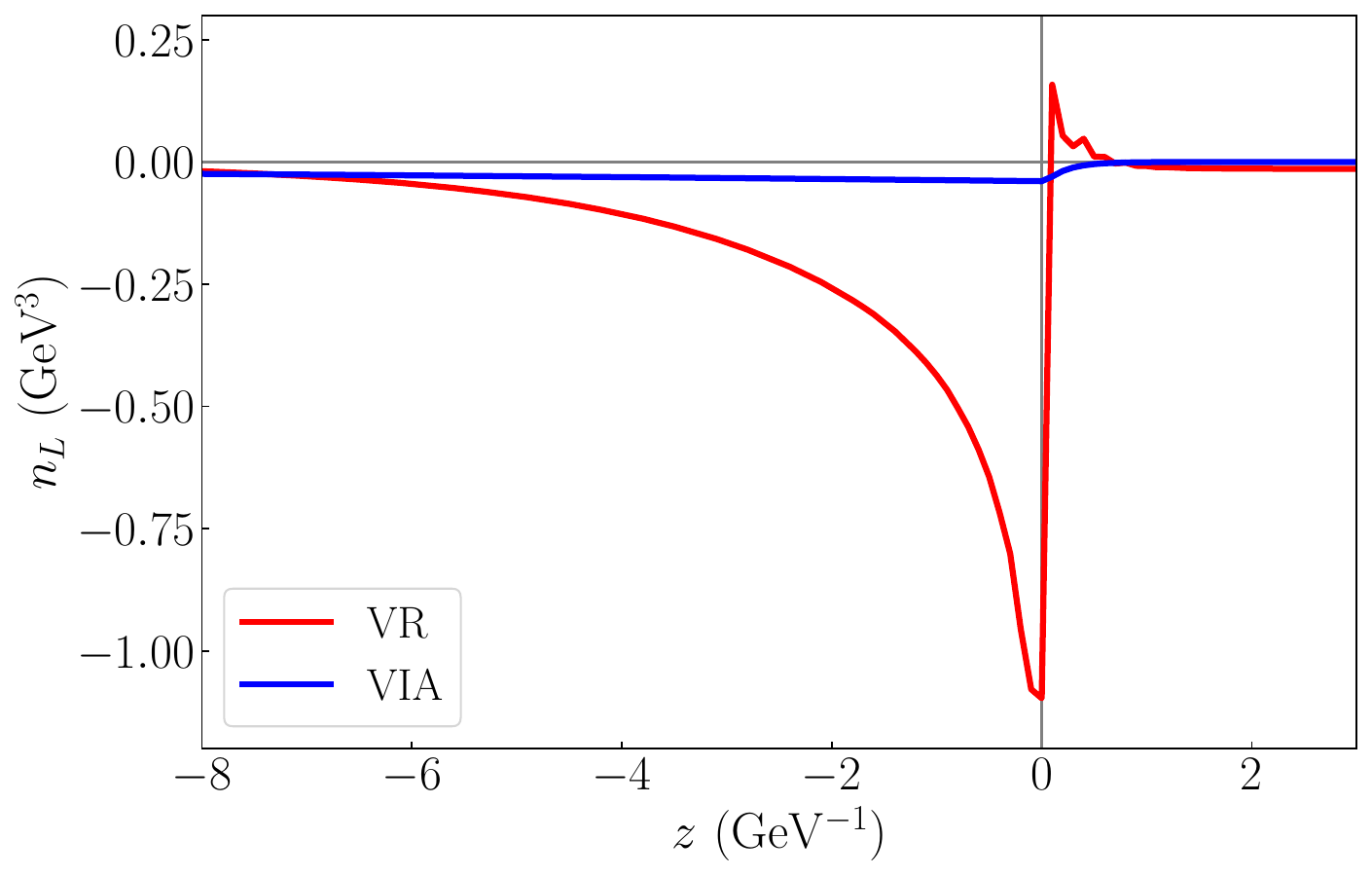}
    \caption{ 
    Left handed quark density $ n_L =  (4 c_T + 5c_Q) n_{H_1} $ for the VIA (blue)  and VR (red) approaches. The bubble exterior (interior) corresponds to $z<0$ ($z>0$).} \label{fig:nH_nL}
\end{figure}

\begin{figure}[ht]
	\centering
	\includegraphics[width = 0.9 \linewidth]{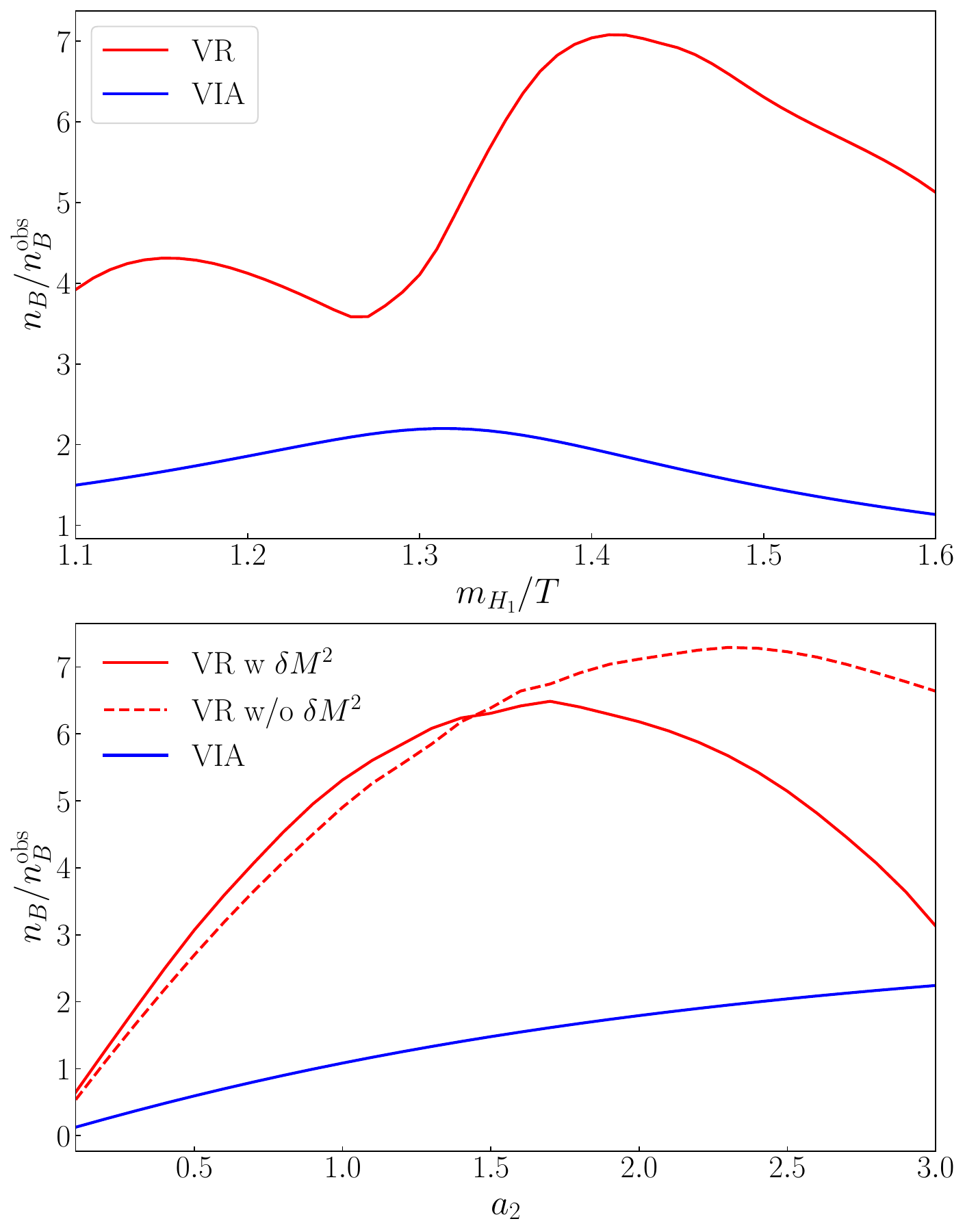}
    \caption{The obtained BAU $ n_B $ as a function of $ m_{H_1} $ (with fixed $ m_{H_2} = 1.32 T $) (top) and the portal coupling $ a_2 $ (bottom) for VR and VIA approaches.}\label{fig:nB_a2}
\end{figure}

This expectation is born out as illustrated in Fig.~\ref{fig:nB_a2} (top), where we show the value of $n_B$ as a function of $m_{H_1}$. 
For both the VR and VIA, the increase in $n_B$ for $m_{H_1}$ near $m_{H_2}$ reflects the resonant enhancement as discussed in Refs.~\cite{Lee:2004we,Cirigliano:2009yt,Cirigliano:2011di}. At the maximum, the VR asymmetry is more than four times larger in magnitude than the VIA value. The double peak structure of VR arises due to a vanishing of the CPV sources $[u\cdot\Sigma, f_m]$ and $[u\cdot\Sigma, \bar{f}_m]$ for $m_{H1}=m_{H2}$\cite{Cirigliano:2009yt}.
Thermal mass corrections induce a slight shift the location of the dip minimum. 

{ The pronounced enhancement of $n_B$ near $m_{H_1} \approx m_{H_2}$ is driven by resonant flavor oscillations. The physical viability of this resonance relies on the interplay between the oscillation frequency, $\Delta\omega_k \equiv |\omega_{1k} - \omega_{2k}|$, and the decoherence induced by plasma interactions. While our VR calculations evaluate the full, coupled integro-differential equations without relying on a simplified damping approximation, we can extract an angle-averaged effective damping rate, $\Gamma_{\text{eff}}(k)$, from the linearized off-diagonal collision integrals (see the Supplementary Material for details). Evaluating this rate at the exact location of the resonance peak for a representative momentum $k = T$, we find $\Delta\omega_k / \Gamma_{\text{eff}}(k) \approx 2.1$, indicating that the quasiparticle and density-matrix formulations employed in this work remain robust, and off-shell effects are not expected to parametrically alter the predicted BAU.}
The VR/VIA enhancement away from this degeneracy point is surprising, as earlier work had suggested the VIA significantly over-estimated the asymmetry. 

Figure~\ref{fig:nB_a2} (bottom) gives the dependence of $n_B$ on the flavor non-diagonal portal coupling $a_2$, illustrating the impact of flavor non-diagonal interactions that enter the VR treatment via the CPV source and CP-conserving collision term. The VIA includes only the former.
Na\"ively, one might anticipate increasing $|a_2|$ would lead to a monontonically increasing $n_B$, owing to correspondingly stronger CPV sources. This expectation is consistent with the VIA curve (blue). In the VR approach, however, for sufficiently large $|a_2|$ the asymmetry begins to decrease, even though the magnitudes of the CPV sources continue to grow. This decrease results from increasingly important damping effects from the CP-conserving collision terms, resulting in closer alignment of the $H_{1,2}$ number densities. An additional suppression at large $a_2$ arises due to flavor non-diagonal thermal mass corrections in the symmetric phase, $\delta M^2$ (dashed red curve). Clearly, a realistic asymmetry computation requires full inclusion and consistent treatment of the CP-conserving interactions, as facilitated by the VR framework.

\begin{figure}[ht]
	\centering
	\includegraphics[width = 0.9 \linewidth]{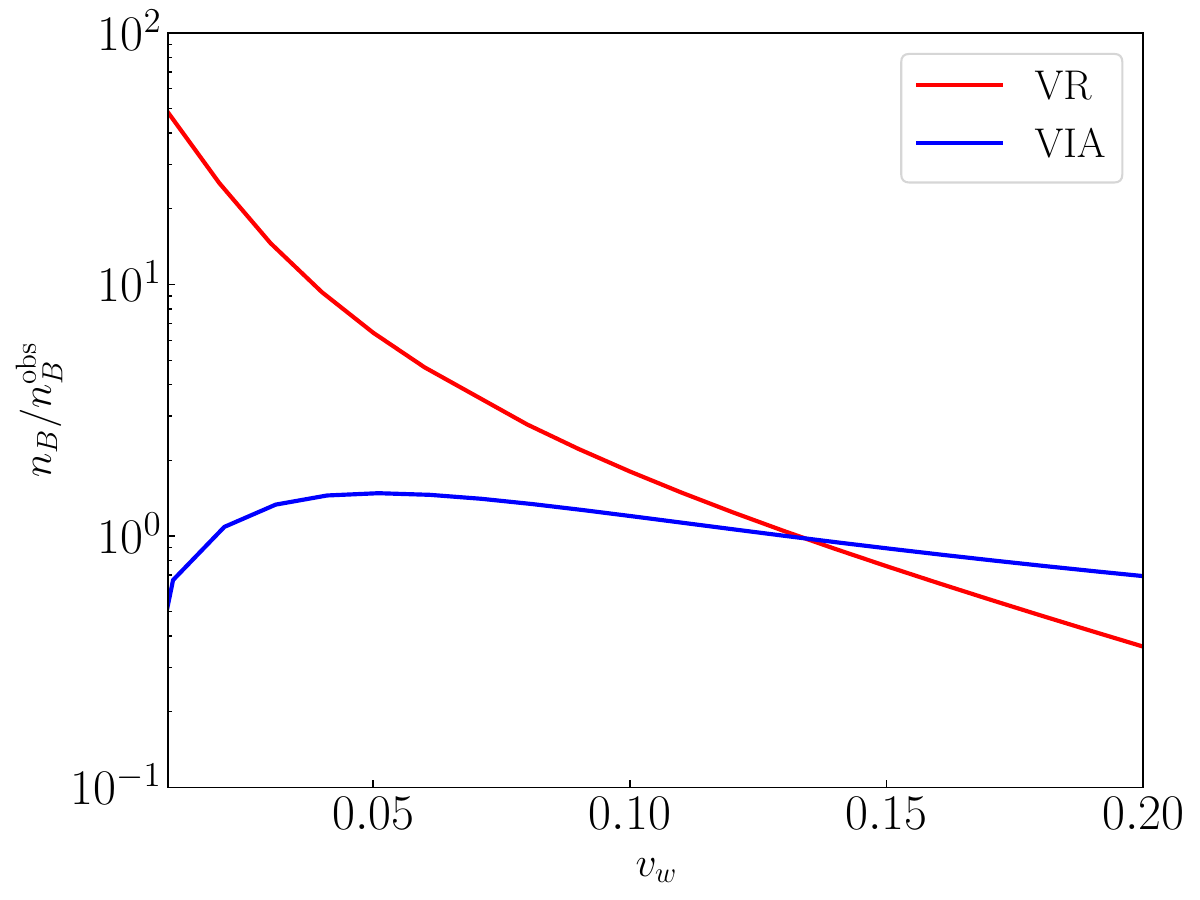}
    \caption{The obtained BAU $ n_B $ as a function of the wall velocity $ v_w $ for VR and VIA approaches.}\label{fig:nB_vw}
\end{figure}

{ One can also study the dependence of the resulting BAU on other model parameters, especially the wall velocity $v_w$, as it affects the BAU through both the particle diffusion and the CP-violating sources. Figure \ref{fig:nB_vw} shows the dependence of the BAU on $v_w$ for both the VR (red) and VIA (blue) approaches, keeping all other benchmark parameters fixed.
{ Typically, the fast-moving wall rapidly sweeps the generated chiral asymmetry into the broken phase, simultaneously decreases the local density $n_L(0)$ and the size of the diffusion tail of $n_L(z)$ ahead of the wall, leaving less time and volume for sphaleron transitions to process the chiral asymmetry. Therefore, increasing $v_w$ generally leads to a smaller BAU. }
Since effects from the CP-conserving collisions and flavor non-diagonal thermal mass corrections are both neglected in the VIA approach, the BAU predicted by {the VIA BAU decreases more slowly with increasing $v_w$ than does the VR BAU.} 
Consequently, the VR approach predicts a significantly larger BAU than the VIA for slow-to-moderate wall velocities ($v_w \lesssim 0.13$), while VIA results may exceed VR results for larger velocities.

In addition, the VIA and VR approaches show distinct behaviors in the slow-wall regime, which is caused by the different wall-velocity dependence of the CPV sources. In the VIA approach, the CPV source is typically proportional to the wall velocity. Consequently, the source vanishes as $v_w \to 0$, causing the sharply dropped BAU. In contrast, the CPV source in the VR framework comes directly from the quantum commutators $[u \cdot \Sigma, f_m]$, which is proportional to the relative velocity of the particles crossing the wall, $v_{\text{rel}} = k_z/\omega_k + v_w$. Therefore, the CPV source in the VR approach remains finite as $v_w \to 0$, resulting in a monotonic increasing diffusion tail $n_L(z)$.}
{ Furthermore, as $v_w \to 0$, the exponential sphaleron washout factor in Eq.~\ref{eq:nB} decays infinitely fast for $z < 0$. Consequently, the $1/v_w$ prefactor perfectly cancels with the integral of the washout exponent, and the resulting BAU is strictly dominated by the chiral asymmetry density at the wall boundary, $n_L(0)$. Therefore, the combination of $v_w$ cancellation and a finite CPV source yields the monotonic increase of the BAU at very low velocities in the VR approach. 

However, we note that Figure \ref{fig:nB_vw} is plotted starting at $v_w = 0.01$ to reflect the physically valid regime of the EWBG transport framework for both the VR and VIA approaches. Specifically, Eq.~\ref{eq:nB} intrinsically assumes that the planar bubble wall acts as a sweeping front that eventually overtakes the entire unbroken phase, safely depositing the generated asymmetry into the broken phase where sphalerons are quenched. 
Mathematically, this sweeping motion is encoded in the lower integration limit of $-\infty$ in Eq.~\ref{eq:nB}, which can be seen as the total spatial reach of the plasma overtaken by the advancing wall.
In the extreme limit of an ultra-slow or stationary bubble wall ($v_w \to 0$), this planar wall sweeping approximation physically breaks down, meaning that the swept reach is simply zero and the integration range should collapse to $[0, 0]$. As a result, a stasis wall would eventually leave all the generated asymmetries in the unbroken phase to be washed out, leading to a vanishing BAU.}

\begin{figure}[ht]
	\centering
       \includegraphics[width = 0.9 \linewidth]{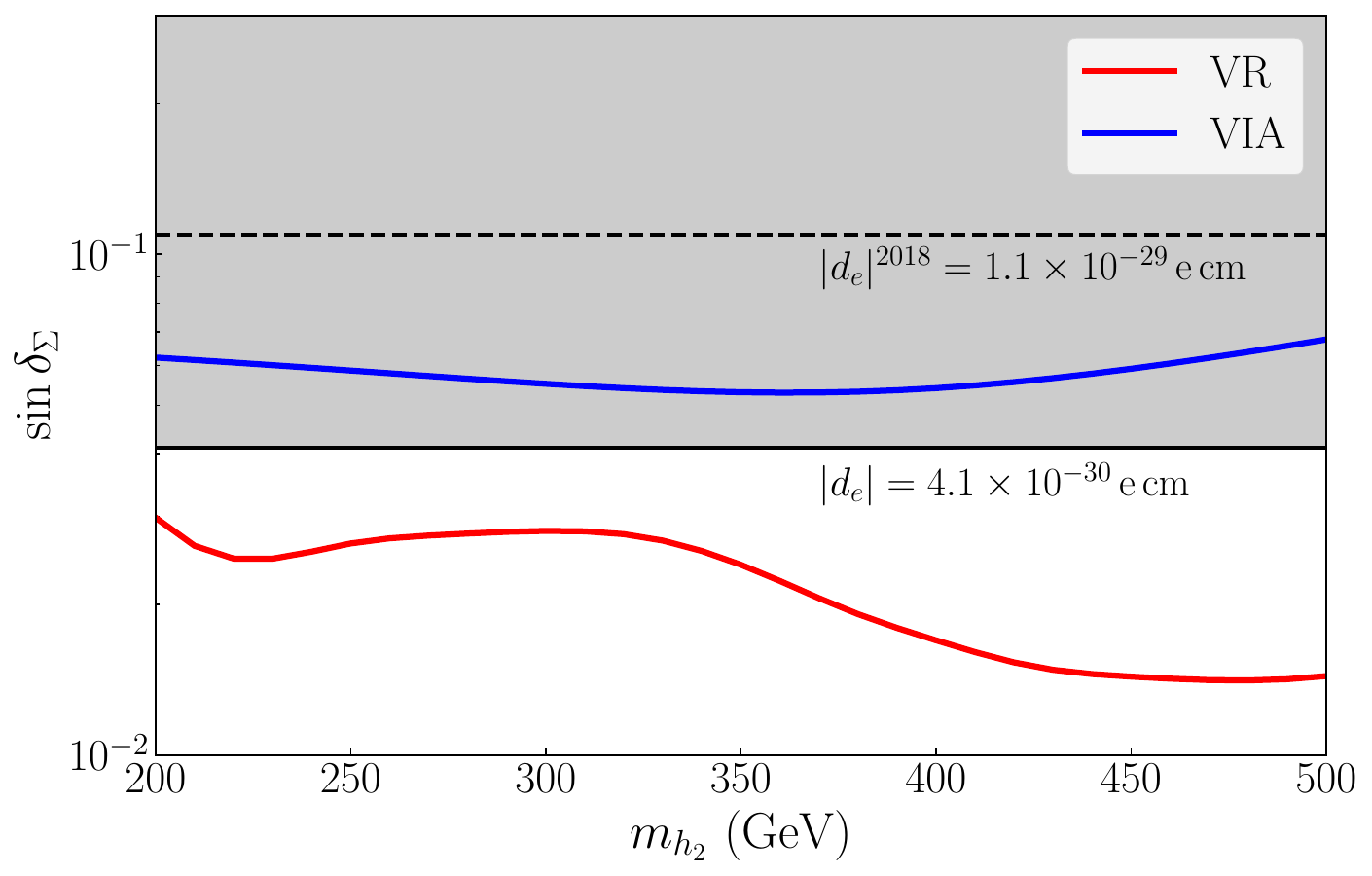}\\
   \caption{Constraints on the CPV phase $\delta_\Sigma$ as a function of the physical $T=0$ mass $m_{h_2}$ with the other parameters fixed. The solid red (blue) band gives the VR (VIA) prediction. The shaded region above the solid (dashed) black line is excluded by the current (previous) electron EDM limit \cite{Alarcon:2022ero} (\cite{ACME:2018yjb}). 
   } \label{fig:EDM2}
\end{figure}

{In Fig.~\ref{fig:EDM2} we show the CPV phase $\delta_\Sigma$ required to generate the observed BAU as a function of $m_{h_2}$, the physical mass of $H_2$ at $T=0$, and compare with the corresponding constraints from experimental limits on $d_e$.}
The latter arises in this model from the two-loop \lq\lq Barr-Zee\rq\rq\ graphs~\cite{Inoue:2015pza}.{{We note that} recent studies of general CP-violating Two-Higgs-Doublet Models have demonstrated that alternative two-loop topologies, known as kite diagrams, can yield non-negligible contributions to the electron EDM (see e.g., Ref.~\cite{Altmannshofer:2020shb}). However, in our specific Two-Step EWBG scenario, these kite contributions are analytically zero. 
Since we assume a CP-conserving 2HDM sector, standard 2HDM kite diagrams contain no CP-violating phase and sum to exactly zero. In addition, the new CP-violating scalar fields ($\Sigma, S$) in our scenario lack Yukawa couplings and possess vanishing vacuum expectation values at $T=0$, therefore they do not mix with the SM-like Higgs doublets and cannot independently form kite topologies. Therefore, the Barr-Zee diagrams evaluated here capture the leading non-vanishing CP-violating effects.}
 
 {Having made these observations, we find that } the present bound $\abs{d_e}<4.1 \times 10^{-30} e \, {\rm cm} $ excludes the shaded region above the solid black line. 
 For reference, we also show the previous $d_e$ bound (dashed black line).  The VR and VIA BAU results are indicated by the red and blue lines, respectively. Importantly, according to the VR computation, this EWBG source remains viable even in light of the new $d_e$ bound. In contrast, the VIA computation -- and by inference the alternative SC approach -- would imply that that model is ruled out.

 We {stress} that application of the VR formulation to other models with {either scalar or fermion} CPV sources {should} also yield more relaxed EDM constraints on EWBG than would be inferred from SC and even VIA treatments, {thereby enhancing the model's EWBG viability}. {Indeed, the specific model used herein illustrates the general features of the first-order-in-gradients CPV sources that will apply to any scenario involving flavor mixing between fields interacting with the bubble wall. For fermion sources, the added complexity of spin necessitates inclusion of spin-dependent particle and antiparticle distribution functions; it will not alter the basic feature of a first-order-in-gradients splitting in the corresponding oscillation frequencies. } Additionally,the state-of-the art treatment of collision, damping, and flavor oscillation dynamics (both thermal and non-thermal) embodied by the VR framework provides the most realistic treatment of these plasma dynamics achieved to date. {This facilitates the resolution of previous inconsistencies between different EWBG CPV dynamic approaches, making the VR framework a realistic approach for confrontation between experiment and any model-specific realization of EWBG.}


\begin{acknowledgements}
We thank V. Cirigliano, {M. Drewes,} K. Kainulainen, and K. Ning for helpful discussions. MJRM was supported in part under U.S. Department of Energy contract DE-SC0011095 {and the National Natural Science Foundation of China under Grants No. 12375094 and W2441004}. JHY and YZL were supported by the National Science Foundation of China under Grants No. 12347105,
No. 12375099 and No. 12047503, and the National Key Research and Development Program of China
Grant No. 2020YFC2201501, No. 2021YFA0718304.

\end{acknowledgements}

\bibliographystyle{apsrev4-1}
\bibliography{references}

\clearpage
\newpage
\onecolumngrid

\centerline{\large {Supplemental Material for}}
\medskip

{\centerline{\large \bf{Does the Electron EDM Preclude Electroweak Baryogenesis ?}}}
\medskip
{\centerline{Yuan-Zhen Li, Jiang-Hao Yu, Michael J.~Ramsey-Musolf}}
\bigskip
\bigskip

In this Supplemental Material, we provide some relevant details of our work, including the form of the mass matrix with the contributions from the vevs and thermal mass corrections, the collision terms in the quantum transport equations, and the benchmark parameters we used in obtaining the figures shown in the main text.

\section{The Mass squared matrix}
Using the Two-Step EWBG model as an example, we consider the interaction potential
\begin{align}
    V_{H\phi} &\supset \frac{1}{2} H_1^\dag H_2\left(a_1 S^2 + a_2 \Sigma^2\right)+ \mathrm{h.c.}
+ \sum_{i=1,2}\left[y_1^{ii}S^2+y_2^{ii}\Sigma^2+y_3^{ii}A^2\right]H_i^\dag H_i\ \ \ , \nonumber \\
&\equiv \frac{1}{2} \left( S^2 \eta^\dagger \, y_S \, \eta + \Sigma^2 \eta^\dagger \, y_\Sigma \, \eta +  A^2 \eta^\dagger \, y_A \, \eta \right),
\label{eq:portalint}
\end{align}
where $\eta\equiv (H^0_1, H^0_2)$ and $y_S, y_\Sigma, y_A$ are the corresponding coupling matrices. {While} $y_A$ is diagonal, $y_S$ and $y_\Sigma$ {contain off-diagonal elements given by} the couplings $a_1$ and $a_2$, respectively. {During the first step of EWSB the mass-squared matrix for the doublet fields, $M_\eta^2$,  depends on the spacetime variation of the $S$ and $\Sigma$ vevs, $v_s$ and $v_\sigma$, respectively, and the couplings in Eq.~(\ref{eq:portalint}). }

{After the second step of EWSB, one has $v_s\to 0$ and $v_\sigma\to 0$, while at $T=0$, the mass-squared matrix can be expressed in terms of the physical masses and mixing angles. Thus,  }
in general, $ M^2_\eta$ can be written as
\begin{equation}
    M^2_\eta(x) = M^2_0 + \Delta M^2_T + \frac{T^2}{24} \left( y_S + y_\Sigma+ y_A \right) + \frac{1}{2} \left( y_S v_S^2(x) + y_\Sigma v_\Sigma^2(x) \right) \,,
\end{equation}
where $M_0^2$ {can be expressed in terms of the $T=0$ phenomenological parameters; $\Delta M^2_T$ gives} the thermal mass contribution from all the other flavor-diagonal thermalizing interactions in the plasma; the third term represents the first order thermal mass correction induced by the one loop self-energy graphs {of} { $V_{H\phi}$}, as seen in FIG.~\ref{fig:self_energy}; and the last term represents the contribution from the vevs $v_S(x)$ and $v_\Sigma(x)$.

{During the first step of EWSB, and after} collecting all the diagonal and non-diagonal contributions, {one may parameterize} $ M^2_\eta$ as 
\begin{equation}
M^2_\eta(x) = \left( 
\begin{array}{cc} 
M_1^2(x) & R(x) \, e^{-i \, \alpha(x)} \\ R(x) \, e^{i \, \alpha(x)} & M_2^2(x) 
\end{array}
\right) \; , \label{eq:wow_appendix}
\end{equation}
where $R(x)$ and $\alpha(x)$ are functions of the CPV phases and vevs,
\begin{align}
    R(x) &\equiv \sqrt{A^2 + B^2}, \quad \alpha(x) \equiv \frac12 \arctan\left(\frac{B(x)}{A(x)}\right), \\
    A(x) &\equiv \frac12 \left( \abs{a_\Sigma} \cos \delta_\Sigma v_\Sigma^2(x) + \abs{a_S} \cos \delta_S v_S^2(x)   \right) + \frac{T^2}{24} \left( \abs{a_\Sigma} \cos \delta_\Sigma  + \abs{a_S} \cos \delta_S  \right), \\
    \quad B(x) & \equiv \frac12 \left( \abs{a_\Sigma} \sin \delta_\Sigma v_\Sigma^2(x) + \abs{a_S} \sin \delta_S v_S^2(x)  \right)+ \frac{T^2}{24} \left( \abs{a_\Sigma} \sin \delta_\Sigma  + \abs{a_S} \sin \delta_S  \right) .
\end{align}
The violation of the CP invariance of the theory requires a spacetime dependent phase of the mass squared matrix, i.e., $\partial_\mu \alpha (x) \ne 0$, so one needs at least two scalar fields to get vevs during the phase transition.

We further diagonalize $M_\eta^2(x)$ at each spacetime point using a unitarity transformation ${\hat \eta}=U(x) {\eta}$, so that $U(x)$ take the form as
\begin{equation}
    U(x) = \begin{pmatrix} \cos\theta(x) & -\sin\theta(x)\, e^{-i \alpha(x)} \\ \sin\theta(x)\, e^{i \alpha(x)} & \cos\theta(x) \end{pmatrix}, \,\, \tan(2 \theta) = \frac{2 R(x)}{M_1^2(x) - M_2^2(x)} \,.
\end{equation}
Note that the mixing angle $\theta$ and phase $\alpha$ depend on the CPV phases through both the contribution of vevs and thermal mass corrections. Our results show that the latter play a significant role in the accurate prediction of the BAU, while they are usually overlooked in previous studies.

\section{The collision term}
\subsection{Collision terms in the quantum Boltzmann equations}

\begin{figure}
    \centering
    \includegraphics[width=0.9\linewidth]{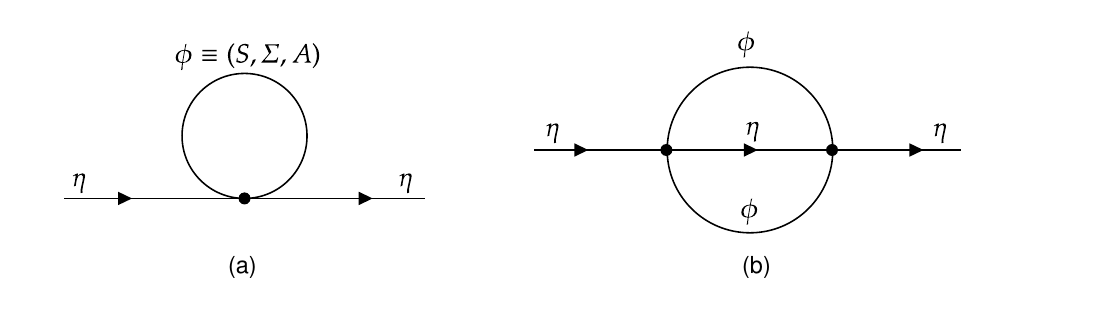}
    \caption{Leading-order graphs contributing to the self-energy terms in the Boltzmann equations, corresponding to (a) thermal mass correction from coherent forward scattering, and (b) non-forward scattering ($\eta\phi \leftrightarrow \eta \phi$) and annihilation ($\eta \eta^\dagger \leftrightarrow \phi \phi$).}
    \label{fig:self_energy}
\end{figure}

Using the vev resummation (VR) framework developed in Refs.~\cite{Cirigliano:2009yt,Cirigliano:2011di}, the quantum Boltzmann equations for the density matrices of the scalar field are derived from the Kadanoff-Baym transport equations \cite{Schwinger:1960qe, Mahanthappa:1962ex, Bakshi:1962dv, Bakshi:1963bn, Keldysh:1964ud,Chou:1984es}, and can be written as
\begin{subequations}
\begin{align}
(u\cdot \partial_X + {\vec F}\cdot\nabla_k ) f_m({\vec k},X) & = -\left[ i\omega_k +u\cdot\Sigma, f_m({\vec k},X)\right]
+\mathcal{C}_m[f_m, {\bar f_m}]({\vec k},X) \ \ \ , \\
(u\cdot \partial_X + {\vec F}\cdot\nabla_k ) {\bar f}_m({\vec k},X) & = +\left[ i\omega_k - u\cdot\Sigma, {\bar f}_m({\vec k},X)\right]
+\mathcal{C}_m[ {\bar f_m}, f_m]({\vec k},X)  \ \ \ ,
\end{align}
\end{subequations}
where $f_m$ and $\bar{f}_m$ are the density matrix for the scalar fields $\eta\equiv (H^0_1, H^0_2)$,  and $u^\mu=(1,{\vec v}) ; \, {\vec v} = {\vec k}/{\bar\omega}_k $; $ {\bar\omega}_k  = \sqrt{|{\vec k}|^2+{\bar m}^2(x)} $; ${\bar m}^2=(M_1^2+M^2_2)/2 ; \; {\vec F}  = \nabla_X {\bar\omega}_k$;  {$\omega_k = {\rm diag} \{ \omega_{1k}, \omega_{2k}\}$; $\omega_{ik} =  \sqrt{|{\vec k}|^2+{ M_i}^2(x)}$.} In addition, $\Sigma^\mu (x) = U^\dag(x) \partial^\mu U(x)$ is the CP-violating source term, and the \lq\lq collision term" $\mathcal{C}_m$ is given by
\be
\mathcal{C}_m[f_m,\bar f_m] \; \equiv \; \int^\infty_0 \! \frac{dk^0}{2\pi} \, \frac{1}{2} \, \left(  \bigl\{\Pi^>(k,X),G^<(k,X)\bigr\} - \bigl\{\Pi^<(k,X),G^>(k,X) \} \, \right) \label{colldef} ,
\ee
where $G^{\gtrless} (k,X) $ are the scalar filed Wightman functions, and $\Pi^{\gtrless} (k,X)$ is the corresponding self-energy functions. For simplicity, the subscript 'm' is omitted below.

For the aforementioned interaction potential 
\begin{equation}
    V_{H\phi} \supset \frac{1}{2} H_1^\dag H_2\left(a_1 S^2 + a_2 \Sigma^2\right)+ \mathrm{h.c.} + \sum_{i=1,2}\left[y_1^{ii}S^2+y_2^{ii}\Sigma^2+y_3^{ii}A^2\right]H_i^\dag H_i =\sum_{\phi\equiv (S,\Sigma,A)} \eta^\dagger Y_\phi \eta \, ,
\end{equation}
the leading-order self energy graphs are shown in Fig. \ref{fig:self_energy}, where the 1-loop graph in Fig. (\ref{fig:self_energy}a) contributes to the thermal mass correction for the mass squared matrix $M_\eta^2$ from coherent forward scattering between $\phi$ and $\eta$, while the 2-loop graph in Fig. (\ref{fig:self_energy}b) gives the non-forward scattering ($\eta\phi \leftrightarrow \eta \phi$) and annihilation ($\eta \eta^\dagger \leftrightarrow \phi \phi$) terms.
Substituting the self-energy functions of Fig. (\ref{fig:self_energy}b) in the collision term Eq. \eqref{colldef} and projecting out the density matrix $f_m, \bar{f}_m$ from the Wightman functions, the collision term for the scattering process ($\eta (k) \phi(p) \leftrightarrow \eta(k^\prime) \phi(p^\prime) $) is given by
\begin{align}
&\mathcal{C}_\phi [f, \bar f]_{\textrm{scat}} = \; - \; \frac{1}{4 k^0}  \int\!\! \frac{d^3 k^\prime}{(2\pi)^3 2 k^{\prime 0}} \int\!\! \frac{d^3 p}{(2\pi)^3 2\epsilon_{\mathbf p}} \int\!\! \frac{d^3 p^\prime}{(2\pi)^3 2\epsilon_{\mathbf p^\prime}} \; (2\pi)^4 \, \delta^4(k+p-k^\prime - p^\prime) \label{Cscat} \\
&\; \times \left[  \bigl\{ f(\mathbf k) , \, Y_\phi (1 + f(\mathbf k^\prime))Y_\phi  \bigr\}f_\phi(\mathbf p)(1+ f_\phi(\mathbf p^\prime)) -  \bigl\{ (1+f(\mathbf k)), \, Y_\phi f(\mathbf k^\prime) Y_\phi  \bigr\}(1+f_\phi(\mathbf p))f_\phi(\mathbf p^\prime)  \right] \notag ,
\end{align}
and the collision term for the annihilation process ($\eta (k) \eta^\dagger (k^\prime) \leftrightarrow \phi(p) \phi(p^\prime) $) is given by
\begin{align}
&\mathcal{C}_\phi [f, \bar f]_{\textrm{ann}} = \; - \; \frac{1}{8 k^0 } \int\!\! \frac{d^3 k^\prime}{(2\pi)^3 2 k^{\prime 0}} \int\!\! \frac{d^3 p}{(2\pi)^3 2\epsilon_{\mathbf p}} \int\!\! \frac{d^3 p^\prime}{(2\pi)^3 2\epsilon_{\mathbf p^\prime}} \; (2\pi)^4 \, \delta^4(k+k^\prime -p - p^\prime) \label{Cann}\\
& \times \left[  \bigl\{ f(\mathbf k) , \, Y_\phi \bar f(\mathbf k^\prime)Y_\phi  \bigr\}(1+ f_\phi(\mathbf p))(1+ f_\phi(\mathbf p^\prime)) -  \bigl\{ (1+f(\mathbf k)), \, Y_\phi (1+\bar f(\mathbf k^\prime))Y_\phi  \bigr\} f_\phi(\mathbf p) f_\phi(\mathbf p^\prime) \right] \notag   ,
\end{align}
where $f_\phi(\mathbf p) \!=\! n_B(\epsilon_{\mathbf p}) \equiv 1/(\exp(\epsilon_{\mathbf p}/T) -1 )$ is the distribution function of the $\phi$ bosons.
In this way, the collision term embodies the effect of all interactions that lead to thermalization in the plasma, chemical equilibrium associated with particle species changing reactions, and diffusion ahead of the advancing bubble wall. 
In addition, the dynamics of the $A$ filed represents all other flavor-diagonal thermalizing interactions in the plasma, such as those arising from gauge and Yukawa interactions. Considering the degrees of freedom in the plasma during the EWPT are around $g_\ast \sim 200$ , the total collision term can be written as
\begin{equation}
    \mathcal{C}[f, {\bar f}] = g_\ast \left( \mathcal{C}_A [f, \bar f]_{\textrm{scat}} + \mathcal{C}_A [f, \bar f]_{\textrm{ann}} \right) + \sum_{\phi = (S,\Sigma)} \left( \mathcal{C}_\phi [f, \bar f]_{\textrm{scat}} + \mathcal{C}_\phi [f, \bar f]_{\textrm{ann}} \right).
\end{equation}
Note that although the diagonal part of the total collision term is dominated by the collision term of $A$, the non-zero off-diagonal contributions from the collision terms of $\Sigma$ and $S$ in the unbroken phase also play a significant role in the diffusion ahead of the advancing bubble wall. Similar to the off-diagonal contributions to the thermal mass correction, the effect of such off-diagonal collisions are also usually neglected in previous studies.
Although the analysis in this letter is based on the 2-Step EWBG model, the VR approach we used can be applied more generally to other EWBG scenarios.

\subsection{Approximations for quantum Boltzmann equations}
In this subsection we present some further approximations or simplifications for the quantum Boltzmann equations (\ref{eq:qb1}, \ref{eq:qb2}) following Refs.~\cite{Cirigliano:2009yt,Cirigliano:2011di}, which are useful for understanding the treatment of the collision terms.

First, we consider EWBG in a late time regime compared to the bubble nucleation time, so that the spherical bubble can be effectively treated as a planar wall, where $z<0$ ($z>0$) corresponds to the unbroken (broken) phase. Therefore, the space-time dependence of the relevant quantities is only presented via the comoving coordinate $z = X+ v_w t$, the distance to the wall,  with $v_w$ the wall velocity.
In addition, we choose the couplings $y_a^{ii}$ ($a=1,2$, $i=1,2$) so as to yield ${\bar m}^2$ to be $x$-independent, implying that ${\vec F}=0$ in our set up.
Consequently, the quantum Boltzmann equations become
\begin{subequations}
\begin{align}
v_{\textrm{rel}} \partial_z  f_m({\vec k},z)  = & -\left[ i\omega_k (z) + v_{\textrm{rel}}\Sigma (z), f_m({\vec k},z)\right]+\mathcal{C}_m[f_m, {\bar f_m}]({\vec k},z)\label{eq:QBE1}\\
v_{\textrm{rel}} \partial_z  {\bar f}_m({\vec k},z)  = & +\left[ i\omega_k (z) - v_{\textrm{rel}}\Sigma (z), {\bar f}_m({\vec k},z)\right] +\mathcal{C}_m[ {\bar f_m}, f_m]({\vec k},z) \label{eq:QBE2}
\end{align}
\end{subequations}
where $v_{\textrm{rel}} (\vec k) \equiv \vec k \cdot \vec n /\bar \omega_k + v_w$ is the relative velocity with respect to the wall, in which $\vec n$ is the unit vector normal to the bubble wall.

In addition, the boundary condition for the quantum Boltzmann equations is that, far from the wall the density matrix reach equilibrium:
\be \label{eq:boundary}
\lim_{z \to \pm \infty} f(\mathbf k,z) , \, \bar{f}(\mathbf k,z) = \lim_{z \to \pm \infty} f^{\textrm{eq}}(\mathbf k,z) \, , \quad
f^{\textrm{eq}}(\mathbf k,z) \equiv  \begin{pmatrix}
    n_B(\omega_{1\mathbf k}(z)) & 0 \\ 0 & n_B(\omega_{2\mathbf k}(z))
\end{pmatrix} \; .
\ee
An important subtlety is then to ensure that the collision term relaxes the density matrix to this "true equilibrium" rather than the "false equilibrium" $f, \bar{f} \!\to\! \textrm{diag} (n_B(\bar{\omega}_{\mathbf k}),n_B(\bar{\omega}_{\mathbf k}))$, where the CP-violating charge generation is unphysically quenched.
To do this, we choose the general solutions of the Wightman functions as
\begin{equation}
\label{eq:treesolutionv2}
\begin{split}
G_{ij}^{>}(k,x)  &= 2\pi\delta(k^2 - {m}_{ij}^2) \, \left[\, \theta(k^0)(\delta_{ij} + f_{ij}(\mathbf k, x)) + \theta(-k^0) \bar {f}_{ij}(-\mathbf k,x) \, \right] ~,  \\
G_{ij}^{<}(k,x)  &= 2\pi\delta(k^2 - {m}_{ij}^2) \, \left[\,\theta(k^0) f_{ij} (\mathbf k,x) + \theta(-k^0)(\delta_{ij}+ \bar f_{ij} (-\mathbf k,x)\, \right] \,.
\end{split}
\end{equation}
with $m_{ij}^2 = 1/2 (m_i^2 + m_j^2)$ and
\begin{equation}
    k^0 = \omega^{ij}_{\mathbf k} \equiv \left\{\begin{array}{lll} \omega^{i}_{\mathbf k} & \quad & i=j \\ \bar{\omega}_{\mathbf k} & \quad & i \ne j \\ \end{array} \right. \; .
\label{massresum}
\end{equation}
By doing this, the collision terms involving diagonal modes are actually evaluated to all orders in $\epsilon_{\rm osc}$, which is the prerequisite for treating equilibration far from the wall properly. 
For the collision terms involving non-diagonal modes, it is efficient to work in the order in $\epsilon_{\rm coll}$, as these modes are damped to zero on a scale $L_{\textrm{mfp}}$ and are not sensitive to longer scales $L_{\textrm{mfp}}/\epsilon_{\rm osc}^n$.

 

Finally, the collision terms for the scattering and annihilation processes can be further simplified for the simplicity of numerical implementation. Eventually, the 9D momentum-integrals in Eq. \eqref{Cscat} and Eq. \eqref{Cann} can be systematically reduced to 3D integrals of the following forms,
\begin{align}
\label{collisionscat}
(\mathcal{C}_\phi)^{\textrm{scat}}_{ij}[f,\bar f] =& \int \!\! \frac{d^3 k^\prime}{(2\pi)^3} \; \left( \delta_{ia} (Y_\phi)_{bc} (Y_\phi)_{dj} + (Y_\phi)_{ic} (Y_\phi)_{da} \delta_{bj} \right) \\
& \times \left( (R_\phi)^{\textrm{scat,in}}_{abcd}(\mathbf k,\mathbf k^\prime) \, (1+f(\mathbf k))_{ab} \, f_{cd}(\mathbf k') - (R_\phi)^{\textrm{scat,out}}_{abcd}(\mathbf k,\mathbf k^\prime) \, f_{ab}(\mathbf k) \, (1+f(\mathbf k'))_{cd} \right) \notag ,\\
\label{collisionann}
(\mathcal{C}_\phi)^{\textrm{ann}}_{ij}[f,\bar f] =& \int \!\! \frac{d^3 k^\prime}{(2\pi)^3} \; \left( \delta_{ia} (Y_\phi)_{bc} (Y_\phi)_{dj} + (Y_\phi)_{ic} (Y_\phi)_{da} \delta_{bj} \right) \\
&\times \left(  (R_\phi)^{\textrm{ann,in}}_{abcd}(\mathbf k,\mathbf k^\prime) \, (1+f(\mathbf k))_{ab} \, (1+\bar f(\mathbf k^\prime))_{cd} - (R_\phi)^{\textrm{ann,out}}_{abcd}(\mathbf k,\mathbf k^\prime) \, f_{ab}(\mathbf k) \, \bar f_{cd}(\mathbf k^\prime) \right)  , \notag
\end{align}
where the scattering kernels are
\begin{align}
{(R_\phi)}^{\textrm{scat,in}}_{abcd}(\mathbf k,\mathbf k^\prime) & = 
\frac{T \, n_B(t_0)}{64\pi t \, \omega_{\mathbf{k}}^{ab} \, \omega_{\mathbf{k}^\prime}^{cd}} \, \theta(t^2-t_0^2)  \, \log\left( \frac{1+n_B(t_-)}{1+n_B(t_+)} \right)  \\
{(R_\phi)}^{\textrm{scat,out}}_{abcd}(\mathbf k,\mathbf k^\prime) & =  \frac{T (1+n_B(t_0))}{64\pi t \, \omega_{\mathbf{k}}^{ab} \, \omega_{\mathbf{k}^\prime}^{cd}} \, \theta(t^2-t_0^2)  \, \log\left( \frac{1+n_B(t_-)}{1+n_B(t_+)} \right) 
\end{align}
and the annihilation kernels are
\begin{align}
{R}^{\textrm{ann,in}}_{abcd}(\mathbf k,\mathbf k^\prime) & =  \frac{T \, n_B(s_0)}{128\pi s \, \omega^{ab}_{\mathbf k} \, \omega^{cd}_{\mathbf k^\prime}} \, \theta(s^2_0-s^2-4m_A^2)  \, \log\left( \frac{n_B(s_-)n_B(-s_-)}{n_B(s_+)n_B(-s_+)} \right)  \\
{R}^{\textrm{ann,out}}_{abcd}(\mathbf k,\mathbf k^\prime) & = \frac{T\, (1+n_B(s_0))}{128\pi s \, \omega^{ab}_{\mathbf k} \, \omega^{cd}_{\mathbf k^\prime}} \, \theta(s^2_0-s^2-4m_A^2)  \, \log\left( \frac{n_B(s_-)n_B(-s_-)}{n_B(s_+)n_B(-s_+)} \right) 
\end{align}
where 
\begin{align}
t &\equiv | \mathbf k - \mathbf k^\prime| , & t_0 &\equiv \omega^{ab}_{\mathbf k} - \omega^{cd}_{\mathbf{k}^\prime} , &
t_\pm &\equiv \, \pm \, \frac{t_0}{2} + \frac{t}{2} \, \sqrt{1+ 4 m_\phi^2/(t^2 - t_0^2) } \; ,  \\
s &\equiv | \mathbf k + \mathbf k^\prime| , & s_0 &\equiv \omega^{ab}_{\mathbf k} + \omega^{cd}_{\mathbf{k}^\prime} , &
s_\pm &\equiv \frac{s_0}{2} \pm \frac{s}{2} \, \sqrt{1+ 4 m_\phi^2/(s^2 - s_0^2) } \; .
\end{align}
The quantum transport equations (\ref{eq:QBE1}, \ref{eq:QBE2}) with the reduced collision terms (\ref{collisionscat}, \ref{collisionann}) for $ \phi\equiv (S,\Sigma,A)$ are then solved using the the \lq\lq relaxation method\rq\rq \cite{Press:1992zz}.

{
In the limit of small deviations from thermal equilibrium, the collision term for the off-diagonal elements of the density matrix, $f_{12} (k,\mu)$, can then be linearized as
\begin{equation}
    \mathcal{C}_{12}[f] \approx -\Gamma_{\text{eff}}(k, \mu) f_{12}(k, \mu) + \text{source terms}.
\end{equation}
where $\Gamma_{\text{eff}}(k, \mu)$ is called as the effective damping rate, quantifying the decoherence effects of the collision terms on the quantum correlation and flavor oscillation.
Using the scattering and annihilation kernels, $\Gamma_{\text{eff}}(k, \mu)$ is given by
\begin{align}
    \Gamma_{\text{eff}}(k, \mu) \approx & \sum_{\phi} \int \frac{d^3k'}{(2\pi)^3}  \left(  (Y_\phi)_{2c} (Y_\phi)_{d2} + (Y_\phi)_{1c} (Y_\phi)_{d1}  \right)  \nonumber \\
    & \times \left( (R_{\phi})_{12cd}^{\text{scat,out}}(k,k')(1+n_B(\omega_{k'})) + (R_{\phi})_{12cd}^{\text{ann,out}}(k,k')n_B(\omega_{k'}) \right)
\end{align}
To benefit the comparison between the resonant oscillation frequency $\Delta\omega_k$, we further define the representative effective damping rate $\Gamma_{\text{eff}}(k)$ by taking the arithmetic average over the angular distribution:
\begin{equation}
    \Gamma_{\text{eff}}(k) = \frac{1}{2} \int_{-1}^{1} d\mu \, \Gamma_{\text{eff}}(k, \mu).
\end{equation}
}

\section{Benchmark Parameters}

{We adopt the following parameter choices in obtaining the figures shown in the main text:} $ a_1=3.0; \, a_2=1.5 ; \, \sin\delta_1=0.1  ; \, \sin\delta_2=0 ; \, y_3^{11}=0.75 ; \, y_3^{22}=1.0$;  {and the thermal masses of $H_1, \, H_2, \, S, \, \Sigma$ and $A$ fields at FOEWPT temperature $T=123 \, {\rm GeV}$ as $m_{H_1}=1.5 T ; \, m_{H_2}=1.32 T ; \, m_s=m_\sigma=0.8 T ; \, m_A = 0.12 \, T$}. {For the VEV Insertion Approximation (VIA) approach, the diffusion constant $D_H$ is also needed to evaluate the baryon asymmetry, for which we choose $D_H=110/T$ as in \cite{Inoue:2015pza}.} 

\end{document}